\documentstyle[aps,psfig,floats,prl]{revtex}
\newcommand {\ignore}[1]{}

\newcommand{\bc}{\begin{center}}
\newcommand{\ec}{\end{center}}

\def\ifmath#1{\relax\ifmmode #1\else $#1$\fi}

\def\3quarter{{\textstyle{3 \over 4}}}

\def\lf{\leaders\hbox to 1em{\hss.\hss}\hfill}
\def\21{$SU(2) \ot U(1)$}
\def\321{$SU(3) \ot SU(2) \ot U(1)$}

\def\mnt{\hbox{$m_{\nu_\tau}$ }}

\def\gau{\hbox{gauge }}

\def\eq#1{{eq. (\ref{#1})}}

\def\lsim{\raise0.3ex\hbox{$\;<$\kern-0.75em\raise-1.1ex\hbox{$\sim\;$}}}
\def\gsim{\raise0.3ex\hbox{$\;>$\kern-0.75em\raise-1.1ex\hbox{$\sim\;$}}}

\def\beq{\begin{equation}}
\def\eeq{\end{equation}}
\def\bef{\begin{figure}}
\def\eef{\end{figure}}
\def\bet{\begin{table}}
\def\eet{\end{table}}
\def\bea{\begin{eqnarray}}
\def\ba{\begin{array}}
\def\ea{\end{array}}
\def\bi{\begin{itemize}}
\def\ei{\end{itemize}}
\def\ben{\begin{enumerate}}
\def\een{\end{enumerate}}

\def\ot{\otimes}
\def\eea{\end{eqnarray}}
\def\np#1#2#3{           {\it Nucl. Phys. }{\bf #1} (19#2) #3}
\def\pl#1#2#3{           {\it Phys. Lett. }{\bf #1} (19#2) #3}
\def\pr#1#2#3{           {\it Phys. Rev. }{\bf #1} (19#2) #3}

\def\n.c.#1#2#3{         {\it Nuovo Cim. }{\bf #1} (19#2) #3}
\def\r.n.c.#1#2#3{       {\it Riv. del Nuovo Cim. }{\bf #1} (19#2) #3}

\def\ppnp#1#2#3{           {\it Prog. Part. Nucl. Phys. }{\bf #1} (19#2) #3}

\twocolumn

\def\gsim{\;\raise0.3ex\hbox{$>$\kern-0.75em\raise-1.1ex\hbox{$\sim$}}\;}
\def\lsim{\;\raise0.3ex\hbox{$<$\kern-0.75em\raise-1.1ex\hbox{$\sim$}}\;}

\begin{document}
\vskip 0.3cm
{\title {\hfill hep-ph/9607401\\
Supersymmetric Unification with Radiative Breaking of R-parity }}
\author{ J. C. Rom\~ao $^\dagger$,
         A. Ioannissyan $^*$ 
and  J. W. F. Valle $^{**}$}
\address{ $^\dagger$ Instituto Superior T\'ecnico, Departamento de F\'{\i}sica\\
A. Rovisco Pais, 1 1096 Lisboa Codex, Portugal; E-mail fromao@alfa.ist.utl.pt}
\address{ $^*$ On leave from Technion, Israel and 
Yerevan Physics Institute, Armenia}
   \address{ $^{**}$ Inst. de F\'{\i}sica Corpuscular, IFIC/CSIC, Dept. de
F\'{\i}sica Te\`orica, Univ. de Valencia, 46100 Valencia, Spain; 
http://neutrinos.uv.es}
\vskip 1cm
\normalsize
\vskip1cm
%

\date{\today}
\maketitle
\begin{abstract}

We show how R-parity can break spontaneously as a result of radiative 
corrections in unified N=1 supergravity models. We illustrate this with 
 a concrete rank-four unified model, where the spontaneous breaking of 
R-parity is accompanied by the existence of a physical majoron. We 
determine the resulting supersymmetric particle mass
spectrum and show that R-parity-breaking signals may 
be detectable at LEP200. 
\end{abstract}


The possible role of supersymmetry in relation
to the hierarchy problem and to the possible
unification of fundamental interactions has
attracted a lot of attention.
Most phenomenological discussions 
have so far been made in the framework  of the Minimal Supersymmetric 
Standard Model (MSSM) \cite{mssm}. Such model assumes a discrete 
symmetry called R-parity \cite{RP}, related to the spin (S), 
lepton number (L), and baryon number (B) according to 
$R_p=(-1)^{(3B+L+2S)}$. Under this symmetry all 
standard model particles are even while their 
partners are odd. Conservation of B and L  
leads to R-parity conservation and implies that SUSY 
particles must always be pair-produced, the lightest 
of them being absolutely stable. 

Whether or not supersymmetry is realized with a conserved 
R-parity is an open dynamical question, sensitive to physics 
at a more fundamental scale. 
On the other hand the phenomenological effects 
associated to R-parity violation may well be 
accessible to experimental verification \cite{beyond}.
It is therefore of great interest to investigate alternative 
scenarios where the effective low energy theory does not
exhibit a conserved R-parity.

As other fundamental symmetries, it could well
be that R-parity is a symmetry at the Lagrangean level 
but is broken by the ground state.  
Such scenarios provide a very {\sl systematic} way to include R 
parity violating effects, automatically consistent with low energy 
{\sl baryon number conservation} and cosmological baryogenesis. 
They may provide an explanation of the of the deficit of solar 
neutrinos and the cosmological dark matter \cite{beyond}.

In this letter we show how R-parity can spontaneously 
break in N=1 supergravity unified  models by virtue of
radiative corrections, very much the same way as the
electroweak symmetry.
We first illustrate how this can happen in the case of rank-four 
unification, such as SU(5), where lepton number is an ungauged
symmetry.
In this case there is a physical Goldstone boson, the Majoron, 
associated to the spontaneous breaking of R-parity. Consistency with 
LEP measurements of the invisible $Z$ width require that R-parity breaking 
be driven by \21 singlet vacuum expectation values (VEVS) 
\cite{MASI,pot3,susymaj}. 
In this case the Majoron is mostly singlet and does not couple to the $Z$.
Here we perform a thorough study of the minimization
of the scalar boson potential and present, as an example, the 
parameters of one of the R-parity-breaking minima we obtain. 
For this minimum we determine the resulting supersymmetric 
particle mass spectrum and show that R-parity-breaking signals 
may be accessible at LEP200.


Starting from some underlying N=1 unified supergravity model 
we consider the low energy theory characterized by the following
\21 invariant superpotential:
\bea
\label{P}
W = h_u u^c Q H_u + h_d d^c Q H_d + h_e e^c \ell H_d + \\ \nonumber
h_0 H_u H_d \Phi +
h_{\nu} \nu^c \ell H_u + 
h \Phi \nu^c S +
\lambda \Phi^3
\eea
The first three terms are the usual ones that will
be responsible for the masses of charged fermions
and the fourth will give rise to the mixing of the
Higgsinos. The last two terms involve gauge singlet
superfields $({\nu^c}, S)$ carrying lepton numbers
-1 and 1, respectively. 
These singlets may arise in several 
extensions of the standard model and may lead to interesting 
phenomenological signatures of their own \cite{fae}.
Their existence ensures that the majoron will
be essentially decoupled from the Z.
The $h_{\nu}$ term plays a crucial role in the
phenomenology, as it will determine the strength 
of the R-parity violating interactions.

All terms in the superpotential in \eq{P} are cubic 
and conserve $total$ lepton number as well as R-parity. 
The superfield $\Phi$ has no lepton number. All couplings 
$h_u,h_d,h_e,h_{\nu},h$ are described by arbitrary matrices 
in generation space but for our present purposes it will be 
enough to assume that they are nonzero only for the third 
generation. We also assume all parameters to be real. 

The model described above is a very simple variant of the one 
proposed in ref. \cite{MASI}.
The matrices $h_d$ and $h_e$ in \eq{P} would be related if we take 
the unification group as SU(5) with minimum Higgs sector.
This relation is not necessary in our analysis and our
results apply also to \321 string models where
the gauge couplings unify by virtue of gravitational 
interactions \cite{Iba}. In this case there are no
relations between the Yukawa matrices.
 
In order to demonstrate the possibility of 
spontaneously breaking R-parity in this model
in a radiative way we write the appropriate
renormalization group equations (RGE) that govern 
the evolution of the parameters. For simplicity
we neglect the $h_\nu$ coupling in the RGE. We will 
neglect, moreover, the bottom-quark Yukawa coupling, 
which is well justified provided $\tan\beta$ is not too large.
First we write the RGE for the Yukawa couplings 
\beq
16 \pi^2 \frac{d h_u}{d t} =
h_u \left( 6 h_u^2 + h_0^2 
- \frac{16}{3} g_3^2 - 3 g_2^2 
- \frac{13}{9} g_1^2 \right) 
\eeq
\beq
16 \pi^2 \frac{d h}{d t} =
h \left( 3 h^2 + 2 h_0^2 + 18 \lambda^2 \right) 
\eeq
\beq
16 \pi^2 \frac{d h_0}{d t} =
h_0 \left( h^2 + 4 h_0^2 
+18 \lambda^2 + 3 h_u^2 
- 3 g_2^2 - g_1^2 
\right) 
\eeq

\beq
16 \pi^2 \frac{d \lambda}{d t} =
\lambda \left( 3 h^2 + 6 h_0^2 + 54 \lambda^2 \right) 
\eeq
where $t = \ln Q/M_U$ where Q is the arbitrary RGE scale
and $M_U$ is the unification scale.
There are similar equations for the evolution of the 
corresponding cubic soft supersymmetry breaking parameters.

The soft-breaking mass parameters evolve according to:
\bea
8 \pi^2 \frac{d M_{H_u}^2}{d t}\hskip -0.2cm  = 
3 h_u^2 ( M_{H_u}^2 + M_Q^2 + M_{u^c}^2 + A_u^2) +  \\\nonumber
h_0^2 ( M_{H_u}^2 + M_{H_d}^2 + M_{\Phi}^2 + A_0^2)  
- 3 g_2^2 M_2^2 - g_1^2 M_1^2
\eea
\bea
8 \pi^2 \frac{d M_{H_d}^2}{d t} =
h_0^2 ( M_{H_u}^2 + M_{H_d}^2 + M_{\Phi}^2 + A_0^2) \\\nonumber
-3 g_2^2 M_2^2 - g_1^2 M_1^2
\eea
\beq
\label{8}
8 \pi^2 \frac{d M_{\nu^c}^2}{d t} = 8 \pi^2 \frac{d M_S^2}{d t} 
= h^2 ( M_{\nu^c}^2 + M_S^2 + M_\Phi^2 + A^2) 
\label{nuc}
\eeq

\bea
8 \pi^2 \frac{d M_\Phi^2}{d t}\hskip -0.2cm = 
2 h_0^2 ( M_{H_u}^2 + M_{H_d}^2 + M_{\Phi}^2 + A_0^2) + \\\nonumber
h^2 ( M_{\nu^c}^2 + M_S^2 + M_\Phi^2 + A^2) 
+18 \lambda^2 ( 3 M_\Phi^2 +  A_\lambda^2 )
\eea

\beq
8 \pi^2 \frac{d M_{\nu_L}^2}{d t} =
-3 g_2^2 M_2^2 - g_1^2 M_1^2
\label{nul}
\eeq

The $g_i$ are the \321 \gau couplings and
the $M_i$ are the corresponding the soft 
breaking gaugino masses. Similarly one can 
write the RGE corresponding to the evolution 
of the soft squark mass terms.

Note that RGE describing the evolution of the $\nu^c$ and $S$
soft supersymmetry breaking masses, given in \eq{8}, are the same
in the limit of negligible $h_\nu$. Moreover, the evolution of the 
stop supersymmetry breaking masses are the same as in the MSSM.

We now discuss the corresponding boundary conditions at unification. 
We assume that at the unification scale the model is characterized 
by one universal soft supersymmetry-breaking mass $m_0$ for all the 
scalars (the gravitino mass), except for the \321 singlets, and a
universal gaugino mass $M_{1/2}$. Moreover we assume that there is 
a single trilinear soft breaking scalar mass parameter $A$. In other 
words we assume that 
\bea
A_u = A = A_0 = A_{\nu} = A_{\lambda} \:, \\
M_{H_u}^2 = M_{H_d}^2 = M_{\nu_L}^2 = M_{u^c}^2 = M_{Q}^2 =m_0^2 \:, \\
M_{\nu^c}^2 = C_{\nu^c}  m_0^2 \ ;M_{S}^2 = C_{S}  m_0^2 \ ;
M_{\Phi}^2 = C_{\Phi}  m_0^2  \:, \\
M_3 = M_2 = M_1 = M_{1/2}
\label{univ}
\eea
at $Q = M_U$. At energies below $M_U$ these conditions do not 
hold, due to the renormalization group evolution from the unification 
scale down to the relevant scale.


In order to determine the values of the Yukawa couplings and 
of  the soft breaking scalar masses at low energies
we first run the RGE from the unification scale 
$M_U \sim 10^{16}$ GeV down to the weak scale. In doing 
this we randomly give values at the unification scale 
for the parameters of the theory. The range of variation 
of these parameters at the unification scale is as follows
\begin{equation}
\begin{array}{ccccc}
10^{-2} & \leq&h^2_t / 4\pi & \leq&1 \cr
10^{-3} & \leq & h^2 / 4\pi ; h^2_0 / 4\pi ;  \lambda^2 / 4\pi & \leq&1 \cr
10^{-7}& \leq&h^2_{\nu} / 4\pi&\leq&1 \cr
-3&\leq&A/m_0&\leq&3 \cr
0&\leq&m_{1/2}/m_0&\leq&2 \cr
\end{array}
\label{unification}
\end{equation}
After running the RGE we have a complete set of parameters, 
Yukawa couplings and soft-breaking masses $m^2_i(RGE)$ to study the minimization. 

The full scalar potential along neutral directions may be
written at low energies as
\beq
V_{total}  = \sum_i \left| { \partial W \over \partial z_i} \right|^2
	+ V_D + V_{SB} + V_{RC}
\label{V}
\eeq
where $z_i$ denotes any one of the neutral scalar fields
in the theory, $V_D$ are the usual $D$-terms, $V_{SB}$ the
SUSY soft breaking terms and $V_{RC}$ are the one-loop radiative
corrections. 

Because of the complexity of the problem we do not do it directly,
solving the non-linear extremization equations for the VEVS. We use,
instead, the procedure developed in \cite{pot3} of solving the extremum 
equations for the soft mass-squared parameters in terms of the VEVS,
which are linear. To do this we have to give values to the VEVS.
We do this in the following way:

\begin{enumerate}

\item
The value of $v_u$ is determined from $m_{top}=h_t v_u$ for
$m_{top}=175 \pm 15$ GeV. If $v_u$ determined in this way is too high 
we go back to the RGE and choose another starting point.

\item
$v_d$ and $\tan(\beta)$ are then determined by $m_W$.

\item
$v_L$ is obtained by solving approximately the corresponding extremum
equation.

\item
We then vary randomly $m_0$, $v_R$, $v_S$, $v_{\phi}$ in the range
$100 \mbox{GeV} \leq m_0 \leq 1000$ GeV and 
$10 \mbox{GeV} \leq v_R ; v_S ; v_{\phi}   \leq 1000$ GeV.

\end{enumerate}

After doing this, for each point in parameter space, we solve the extremum
equations for the soft breaking masses, which we now call $m^2_i$. Then we 
calculate numerically the eigenvalues for the real and imaginary part of 
the neutral scalar mass-squared matrix. If they are all positive, except 
for the Goldstone bosons, the point is a good one. If not, we go to the 
next random value. After doing this we end up with a set of points for which:

\begin{enumerate}

\item
The Yukawa couplings and the gaugino mass terms are given by the RGE.

\item
For a given set of $m^2_i$ each point is also a solution of the minimization
of the potential that breaks R-Parity. 

\item
However,  the $m^2_i$ obtained by the minimization
of the potential differ from those obtained from the RGE $m^2_i(RGE)$. 

\end{enumerate}

Our next goal is to find which solutions for $m^2_i$ that minimize the
effective low-energy potential have the property that they 
coincide with the $m^2_i(RGE)$ obtained, for a given unified theory, 
from the RGE, namely
\begin{equation}
m^2_i=m^2_i(RGE) \quad \forall i
\end{equation}
To do that we define a function
\begin{equation}
\epsilon= max \left( \frac{m^2_i}{m^2_i(RGE)},\frac{m^2_i(RGE)}{m^2_i}
\right) \quad \forall i
\end{equation}

Defined in this way it is easy to see that we have always $\epsilon \geq 1$
the equality being what we are looking for. We are then all set for a
minimization procedure. We want, by varying the parameters, to get
$\epsilon$ as close to 1 as possible. Before we move on we have to clarify
what are our {\it parameters} in the minimization. At first we assumed
universality and our $\epsilon$ was a function
of $h_t^{U}, h^{U}, h_0^{U}, h_{\nu}^{U}$,
$\lambda^{U}, A^{U}, m_0, m_{1/2}, v_R, v_S, v_{\phi}$, 
and the allowed range for these parameters was as 
specified above. 

With these conditions we used the {\tt MINUIT} package to find
the minimum of $\epsilon$. We should add that we have also enforced that we 
get a solution that it is both a solution of the minimization of the 
potential and lower than other trivial minima.
After sampling a few million points we did not find any solution with 
$\epsilon < 1.1$. We then decided to relax the universality condition 
on the soft mass-squared parameters at the unification scale. Indeed,
deviations from universality, are a generic feature of soft-breaking
terms obtained from 4-dimensional string models \cite{brignole}.
For definiteness, we adopted a very conservative and unnecessary 
restriction of keeping universality for the MSSM scalars but allowed 
the \21 singlet masses to vary away from universality. To be more precise 
we defined
\begin{equation}
\eta_S       = \frac{m^2_S}{m_0^2}\quad ; \quad
\eta_{\nu^c} =  \frac{m^2_{\nu^c}}{m_0^2}\quad ; \quad
\eta_{\phi}  =  \frac{m^2_{\phi}}{m_0^2}
\end{equation}
and allowed $\eta_S$, $\eta_{\nu^c}$ and $\eta_{\phi}$ to vary from
$\frac{1}{10}$ to $10$.
Finally we also allowed a variation of the top quark mass within 
present experimental errors.

With these modifications our $\epsilon$  is now a function of
$h_t^{U}, h^{U}, h_0^{U}, h_{\nu}^{U}, \lambda^{U}$,
$A^{U}, m_0, m_{1/2}, v_R, v_S, v_{\phi}, \eta_S, \eta_{\nu^c},
\eta_{\phi}$ and $m_{top}$, and {\tt MINUIT} was able to find 
solutions with $\epsilon$ as close to 1 as we wanted.


Here we present for one specific case the values at the unification 
scale as well as the low energy values and the low energy spectrum. 
The starting values at the unification scale are the following:
\begin{equation}
\begin{array}{ccccc}
&&A =2.99 \:, \\ \nonumber
&&m_0= 143.6\ GeV  \:, \\\nonumber
&&C_{\nu^c}= 0.869 \ ; \ C_{S} = 0.742  \ ; 
C_{\Phi} = 1.204  \:, \\\nonumber
&&M_{1/2} = 0.907\ m_0 \:,\\\nonumber
&&\frac{h_u^2}{4 \pi}= 0.03 \ ; 
\frac{h^2}{4 \pi}= 0.015 \ ; 
\frac{h_{\nu}^2}{4 \pi}= 1.2 \times 10^{-7} \:, \\\nonumber
&&\frac{h_0^2}{4 \pi}= 0.032 \ ; \
\frac{\lambda^2}{4 \pi}= 0.0064 
\label{unificationscale}
\end{array}
\end{equation}
With these values we get the following particle mass spectrum at low scale
\bea
&&m_t=174 \mbox{GeV}; \: \tilde{m}_{t_1}= 295 \mbox{GeV}; \: 
\tilde{m}_{t_2}=435\ \mbox{GeV}, \\
&&m_{\chi^\pm_1}= 78\ \mbox{GeV}; \:  m_{\chi^\pm_2}= 250\ \mbox{GeV}, \\
&&m_{\nu_{\tau}}= 65\  \mbox{KeV}; \:
m_{\chi^0_1}= 43\ \mbox{GeV}; \:
m_{\chi^0_2}= 83\ \mbox{GeV}, \\
&&m_{\chi^0_3}= 221\ \mbox{GeV}; \:
m_{\chi^0_4}= 251 \mbox{GeV}, \\
&&m_{h}=69 \mbox{GeV}; \: m_{H}=161 \mbox{GeV}; \:  m_{A}=198 \mbox{GeV} 
\eea
The shape of the scalar potential close to this minimum
can be displayed as a function of the relevant VEVS, for
example the R-parity violation VEVS $v_R$ and $v_S$ (fig. 1) 
\bef
\centerline{\protect\hbox{
\psfig{file=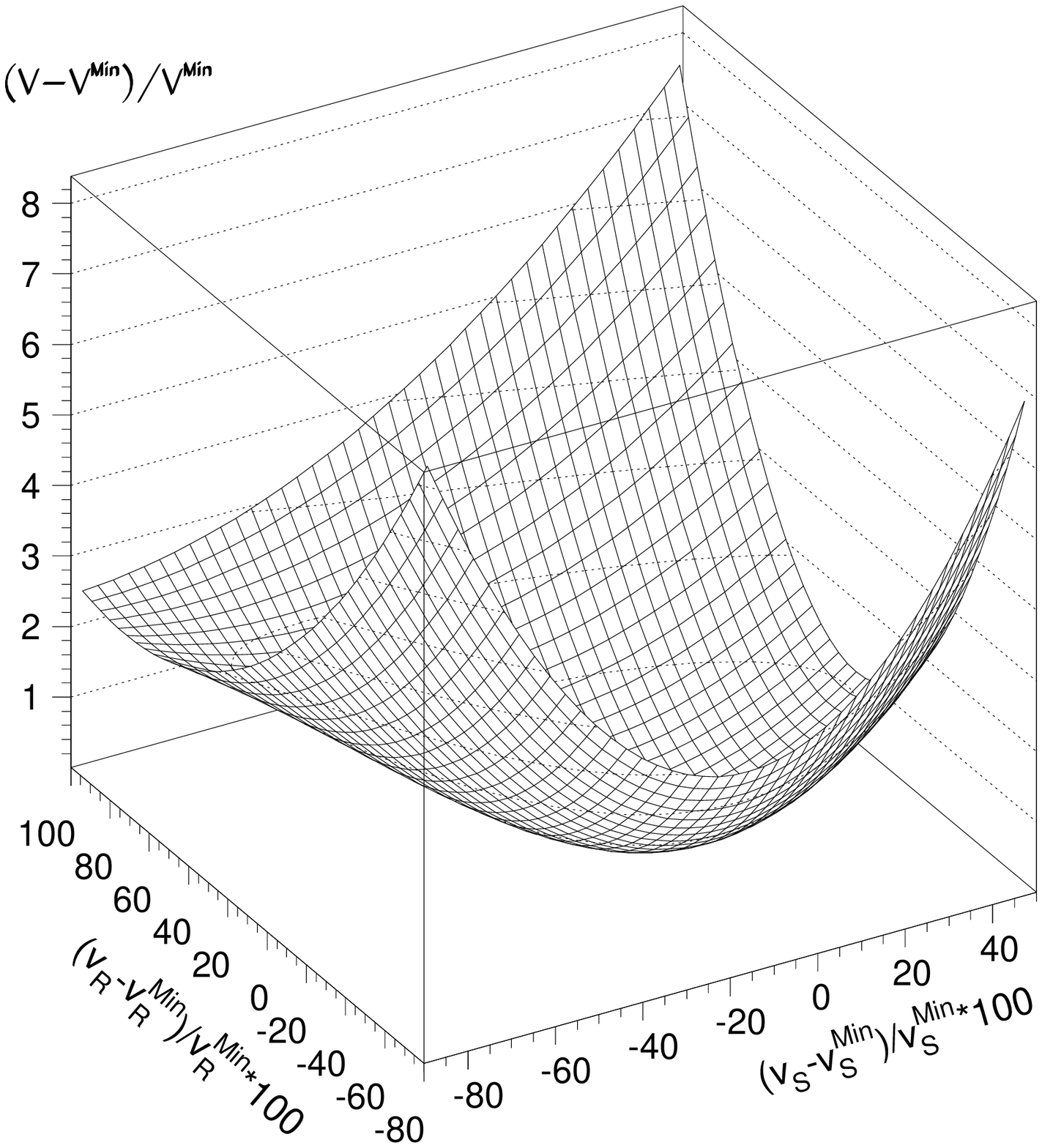,height=6cm}}}
\caption{Shape of the scalar potential close to the minimum
studied in this paper, displayed as a function of the R-parity violation 
VEVS $v_R$ and $v_S$. }
\eef
or the electroweak breaking VEVS $v_u$ and $v_d$. 
We have also checked that the R-parity minimum is lower than trivial
minima, for which electroweak and/or R-parity are unbroken, and
that at all scales the traditional bound for no colour breaking \cite{Frere}
\beq
\label{frerebound}
|A_u|^2 \leq 3\left( m_{Q_u}^2 + m_{u}^2 + m_2^2\right)
\eeq
is satisfied. 


We see that, in this example, the lightest CP-even Higgs boson,
the lightest  chargino and the lightest neutralino can all be 
produced at LEP200. Moreover, since R-parity is broken, the 
lightest neutralino decays. Moreover, typically this decay 
happens in the detector, as can be seen from fig 2. 

 In our model the value of \mnt determines the rates for all 
R-parity-violating couplings. Since the value of \mnt for this 
solution is relatively small (65 KeV), the most likely site for 
the violation of R-parity will be in the decay of the lightest 
neutralino which would arise as the final stage of the cascade 
decays of the other supersymmetric particles. 
Note that the above minimum is just one out of many. There are
others with light SUSY spectra, for which $ m_{\nu_{\tau}}$ lies 
in the tens of MeV range. In the latter case R-parity violation would 
show up not only through the decay of the lightest neutralino, 
but might also be observable at LEP100, e.g. through the single 
production of charginos, as proposed earlier \cite{ROMA}.

\bef
\centerline{\protect\hbox{
\psfig{file=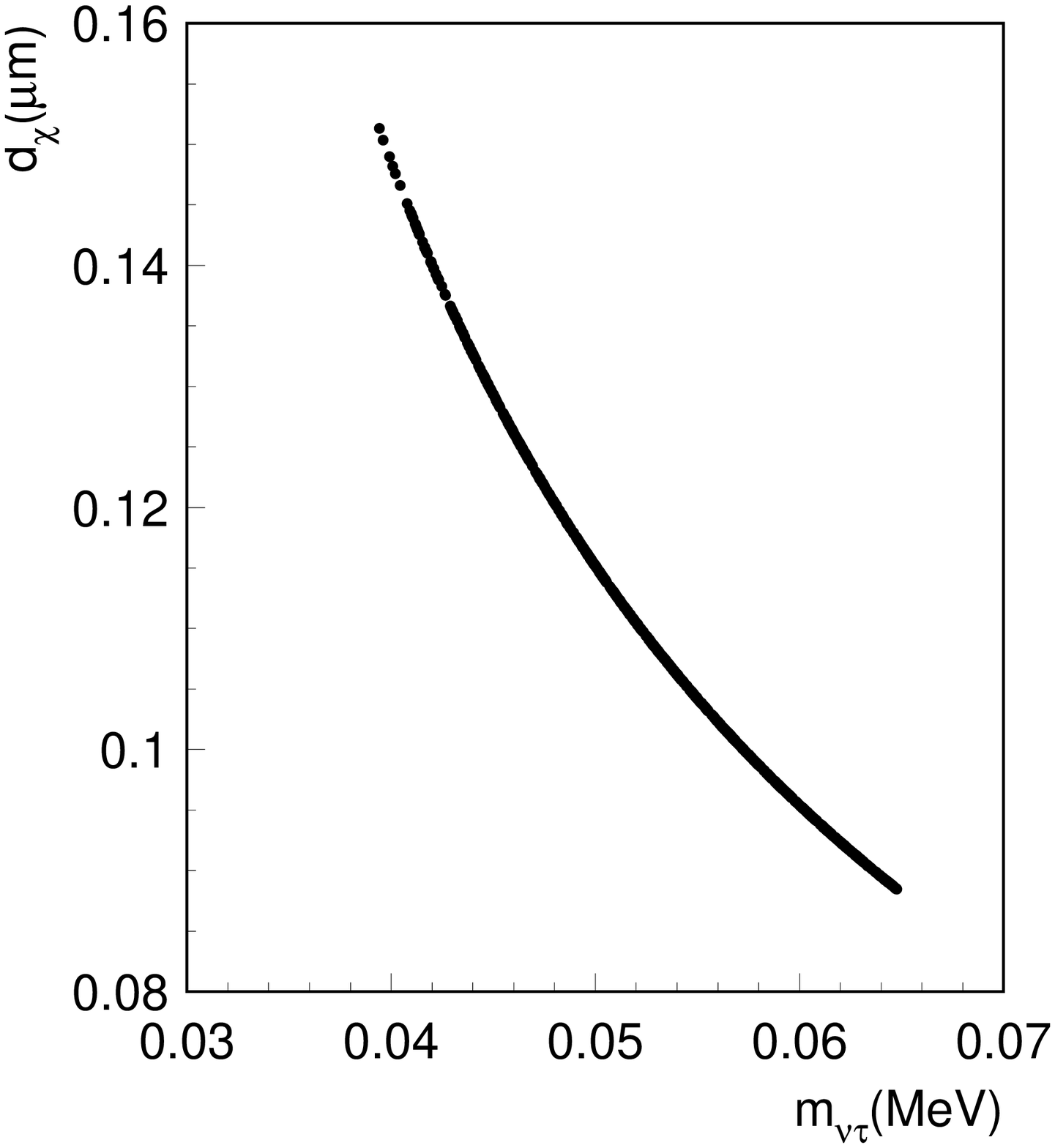,height=6.5cm}}}
\caption{Typical neutralino decay length versus \mnt }
\vglue -.5cm
\eef
Before concluding we wish to comment on the issue of the
universality of soft-breaking masses. The solutions with 
light supersymmetric mass spectrum that we have obtained 
have non-universal values at unification. We do not
know if this is a necessary feature of the model. Were
this to be confirmed by further studies, we would regard it 
as an interesting clue to relate  R-parity-breaking with
physics at the Planck scale in the string context. Indeed,
deviations from universality are a generic feature of 
soft-breaking terms obtained from 4-dimensional strings \cite{brignole}.


This work was supported by DGICYT grants 
PB92-0084 and HP-53 and by a Valencia-Yerphi exchange (A.I.). 
We thank Carlos Mu\~noz for discussions.



\noindent

\bibliographystyle{ansrt}

\end{document}